\documentclass[twocolumn]{revtex4}

\begin{document}

\title{Decaying $\Lambda$ cosmology, varying G and holography}

\author{S. Carneiro$^1$ and J. A. S. Lima$^{2,3}$}

\affiliation{$^1$Instituto de F\'{\i}sica, Universidade Federal da
Bahia, 40210-340, Salvador, BA, Brazil \\ $^2$Departamento de
Astronomia - IAG, Universidade de S\~ao Paulo, 05508-900, S\~ao
Paulo, SP, Brazil \\ $^3$Departamento de F\'{\i}sica, Universidade
Federal do Rio Grande do Norte, 59072-970, Natal, RN, Brazil}

\begin{abstract}
We discuss a class of uniform and isotropic, spatially flat,
decaying $\Lambda$ cosmologies, in the realm of a model where the
gravitation ``constant" $G$ is a function of the cosmological
time. Besides the usual de Sitter solution, the models at late
times are characterized by a constant ratio between the matter and
total energy densities. One of them presents a coasting expansion
where the matter density parameter is $\Omega_m= 1/3$, and the age
of the universe satisfies $Ht=1$. From considerations in line with
the holographic conjecture, it is argued that, among the
non-decelerating solutions, the coasting expansion is the only
acceptable from a thermodynamic viewpoint, and that the time
varying cosmological term must be proportional to $H^2$, a result
earlier obtained using different arguments.
\end{abstract}

\maketitle

\section{Introduction}

One of the most intriguing problems in modern cosmology is the
present observed value of the cosmological constant, which amounts
to nearly 120 orders of magnitude smaller than the value predicted
by quantum field theories \cite{Starobinsky}. A possible
explanation for this huge discrepancy is based on the idea that
the vacuum energy density is not constant, but decays as the
universe expands \cite{OT}-\cite{Shapiro}.

When the vacuum energy density is computed with the help of
quantum field theories in flat spacetime, one obtains a divergent
result, which can be regulated by means of an ultraviolet cutoff
of the order of the Planck mass. In the standard recipe, such a
contribution to the vacuum energy must be exactly canceled by a
bare cosmological constant in the Einstein equations, because, in
the flat spacetime, the right-hand side of those equations is
identically zero \cite{Ralf}. Therefore, in a curved background, a
renormalized value for $\Lambda$ should be obtained by calculating
the vacuum energy density and then subtracting the Minkowskian
result. Since spacetime was strongly curved at Planck times, one
may expect an initial huge value of the order of $l_p^{-2}$, where
$l_p = \sqrt{G}$ is the Planck length. But, as the universe
expands, the observed cosmological constant should decay, thereby
leading to the small value observed nowadays. In other words, the
present value of $\Lambda$ (the vacuum energy density) is small
because the Universe is too old \cite{PR}.

This line of reasoning leads one to infer that other physical
``constants" should also vary at a cosmological scale. Indeed, the
quantum vacuum fluctuations behind the net vacuum energy density
contribute also to the observed, renormalized, values of couplings
and masses of elementary particles. Therefore, if the vacuum
density decays with the universe expansion, one should expect
cosmological time variations of charges and masses. The first
would lead to the variation of the fine structure constant
\cite{Ranada}, an effect that have possibly been observed
\cite{Alfa}. As to the variation of masses, it is equivalent to a
variation of the gravitational constant $G$, at least in what
concerns gravitational effects. The variation of $G$ should also
be expected by considering it a renormalized coupling constant of
a (still unknown) quantum theory of gravity.

Since the seminal papers by Dirac \cite{Dirac}, a possible
variation of $G$ has been investigated with no success by several
teams, through geophysical and astronomical observations, at the
scale of solar system and with binary systems \cite{frances}.
However, it should be stressed that we are talking here about time
variations at a cosmological scale, and cosmological observations
still cannot put strong limits on such a variation, specially at
the late times of the evolution. Naturally, if the $G$ variation
originates from the decaying of vacuum density at a cosmological
scale, one would expect a constant value of $G$ at the scales of
solar system and galaxies, since at these scales the matter
density and the curvature (and presumably the vacuum density) are
nearly constant.

In this paper we discuss a new decaying vacuum scenario which is
supposed to induce a time varying gravitational constant. As we
shall see, the Friedmann equations for the flat case plus the
Eddington-Weinberg relation \cite{GS} imply that beyond the usual
de Sitter solution there are other solutions characterized by a
constant ratio between the matter density and the total energy
density. One of them presents a coasting expansion where the
universe age is fixed by $Ht=1$, and the matter density parameter
is $\Omega_m=1/3$.

The work is structured as follows. Next section we review an
earlier approach to decaying vacuum energy models with varying $G$
\cite{Ademir3,MGX}. In the subsequent sections we make some
considerations in line with the so called holographic principle
\cite{GRF,GS,Bousso} to justify the Eddington-Weinberg relation,
and to argue that, among all the possible non-decelerating
solutions, that with coasting expansion is the only plausible from
a thermodynamic viewpoint.

It should be noticed that models with variation of both $\Lambda$
and $G$ have already been considered in the literature. For
instance, Bertolami \cite{Bertolami} gave a detailed analysis of
varying $\Lambda$ cosmological solutions in the realm of a
Brans-Dicke theory. However, the variation of $G$ we are
interested here has its origin in the variation of the vacuum
energy density. Therefore, although considering that a Brans-Dicke
scalar field might be used as an effective description of that
variation, in what follows we shall make no use of such a field.

\section{Decaying $\Lambda$ solutions}

In the flat case, the Friedman equations with a time dependent $G$
can be written as
\begin{eqnarray}
\rho = \frac{3H^2}{8\pi G}, \\ \dot{\rho}+3H(\rho + p)= 0,
\end{eqnarray}
where $H=\dot{a}/a$ is the Hubble parameter, and a dot means
derivation respect to the cosmological time $t$. The second
equation does not depend on the varying or constant character of
$G$, being just an expression of energy conservation and the
equivalence principle \cite{Ademir3,Aldrovandi}. As far as
relation (1) is concerned, it is valid today, and we will assume
that it is also valid for any time in course of the evolution.

The variation law for $G$ follows from the Eddington-Weinberg
empirical relation \cite{GS}
\begin{equation}\label{EW}
G = \frac{H}{8\pi \lambda} \approx \frac{H}{m^3},
\end{equation}
where $m$ has the order of the pion mass, and the constant
$\lambda$ was introduced for convenience. Once again, this
relation is valid today, and we will assume that it is valid since
early times. Substituting (3) into (1), we have
\begin{equation}
\rho = 3\lambda H.
\end{equation}

Let us now consider a twofold energy content, formed by dust
matter with energy density $\rho_m$ plus a vacuum term with
equation of state $p_{\Lambda}=-\rho_{\Lambda}$. The total energy
density and pressure read
\begin{eqnarray}
\rho &=& \rho_m + \rho_{\Lambda}, \\ p &=& -\rho_{\Lambda}.
\end{eqnarray}
Inserting (4)-(6) into the conservation equation (2), we obtain
\begin{equation}
\lambda \dot{H} + \rho_m H = 0.
\end{equation}

To proceed further, let us now assume that the quantities $H$ and
$\rho_m$, at late times, fall monotonically as power laws of the
scale factor $a$:
\begin{eqnarray}
H &=& \beta/a^k, \\ \rho_m &=& \gamma/a^n,
\end{eqnarray}
where $n$ and $k$ are two positive parameters. Replacing the above
expressions into (7) leads to
\begin{equation}
\gamma a^{-n} - \lambda \beta k a^{-k} = 0.
\end{equation}
The above equation is valid for any (large) value of $a$ only if
\begin{eqnarray}
k&=&n, \\ \gamma &=& \lambda \beta n.
\end{eqnarray}

With these results, and by using (4), (8) and (9), it is easy to
show that
\begin{equation}
\Omega_m \equiv \rho_m/\rho = n/3.
\end{equation}
Since $0 \leq \Omega_m \leq 1$, it follows that $0 \leq n \leq 3$.
In the case $n=0$, we have
\begin{eqnarray}
H&=&\beta, \\ \rho_m&=&0, \\ G&=&\frac{\beta}{8\pi \lambda}.
\end{eqnarray}
This solution corresponds to a de Sitter universe, with $\Lambda$
and $G$ constant.

For the other values of $n$, we obtain from (8) the solution
\cite{geral}
\begin{eqnarray}
a &=& (n\beta t)^{1/n}, \\ H t&=&1/n, \\ q &=& n-1,
\end{eqnarray}
where $q=-a\ddot{a}/\dot{a}^2$ is the deceleration parameter. In
the case $n=1$ we see that $q=0$ and
\begin{eqnarray}
a&=&\beta t, \\ H t &=& 1, \\ \Omega_m &=& 1/3.
\end{eqnarray}

As discussed in \cite{Dev}, among the solutions of the form
$a\propto t^n$, the best fitting of supernova and radio sources
observations is obtained in the coasting case, $a\propto t$. On
the other hand, the relation $Ht=1$ gives the best estimation for
the universe age \cite{age}, and the relative matter density given
by (22) matches surprisingly well the observed value \cite{omega}.

It is worth notice that in a context with varying $\Lambda$ but
constant $G$ the coasting solution with $Ht=1$ corresponds to a
relative matter density equals to $2/3$ \cite{MGX}, in clear
disagreement with observation. To obtain $\Omega_m = 1/3$, we
would have to have $Ht = 2$, completely outside the observed
limits.

Let us determine the variation rate of $G$, as well as the rate of
matter production in this model. The evolution law (3) leads to
the relative variation rate
\begin{equation}
\dot{G}/G = -(1+q)H.
\end{equation}
For $n=1$, we have $q=0$, and so
\begin{equation}
\dot{G}/G = -H.
\end{equation}

The rate of matter production (coming from the decaying vacuum
energy) reads
\begin{equation}
\frac{1}{\rho_m a^3}\frac{d}{dt}(\rho_m a^3)= (3-n) H.
\end{equation}

Let us now discuss the variation of the vacuum energy density in
the present context. By using (3)-(5) and (13), one may check that
the vacuum energy density and the corresponding cosmological
constant are given by
\begin{eqnarray}
\rho_{\Lambda} &=& (3-n)\lambda H \approx
\frac{(3-n)}{8\pi}\;m^3H,
\\ \Lambda &=& 8\pi G \rho_{\Lambda} = (3-n) H^2,
\end{eqnarray}
leading to present values in agreement with observation.

The first above equation is in accordance with a recent derivation
by Sch\"utzhold \cite{Ralf}, based on quantum field estimations in
an expanding background. The second equation agrees with an ansatz
proposed by Chen and Wu on the basis of dimensional arguments
\cite{Wu}, and further modified by Carvalho et al. \cite{Ademir}
in order to obtain the $H^{2}$ dependence. It also matches a
result based on a renormalization approach \cite{Shapiro}, in the
spirit discussed in the Introduction. Note that, in the realm of a
constant $G$ cosmology, those two equations would be incompatible
to each other \cite{Zeldovich}.

\section{Holography and $G$ variation}

Since the initial proposals on the existence of entropy bounds in
self-gravitating systems, rigorous verifications of this idea have
been obtained in some cases, leading to the general belief that
there is an entropy bound associated to any horizon, of the order
of the horizon area (in Planck units) \cite{Bousso}. For our
purposes, this holographic conjecture could be understood as
follows.

Consider a set of degrees of freedom inside a horizon of
characteristic radius $R$. The limitation of its configuration
space leads to the quantization of its energy-momentum space, with
a quantum of energy given by $m_0 \approx 1/R$ \cite{Wesson}.
Then, the maximum number of degrees of freedom inside the horizon
is given by $N \approx M/m_0 \approx MR$, where $M$ is the mass
inside the horizon. So, for any system obeying the relation $M
\approx R/G$, we have $N \approx R^2/G$, as just established by
the holographic conjecture.

The reader knows two important examples of systems satisfying the
above conditions. The first is a black-hole, for which we have (in
the non-rotating case) $M=R/(2G)$, where $R$ is the gravitational
radius. The second is the spatially flat Friedmann universe.
Indeed, in this case the total energy density is given by $\rho =
3/(8\pi G R^2)$, where $R$ is the Hubble radius. On the other
hand, the volume inside the Hubble horizon is $4\pi R^3/3$, which
leads to a total mass inside the Hubble sphere given, again, by
$M=R/(2G)$. Therefore, the holographic conjecture applies as well
to our universe: the number of observable degrees of freedom is
bounded by the area of the Hubble horizon.

But how many degrees of freedom we have in the observable
universe? We know that the number of barions is of the order of
$10^{80}$, and that the Universe contains about $10^{8}$ photons
per barion. It is also reasonable to believe that the contribution
of the non-barionic dark matter may be of the same order of
magnitude.  But the major contribution to the entropy of matter
seems to come from massive black-holes present in galactic nuclei,
which represents an entropy of the order of $10^{101}$
\cite{Penrose}.

As far as the vacuum entropy is concerned, it is not trivial to
define the number of virtual particles in curved backgrounds. Let
us remind, however, that our universe has a quasi-flat spacetime.
Therefore, the following estimation can be considered a good
approximation \cite{GRF}. It is clear that vacuum configurations
of classical fields (as the vacuum expectation value of the Higgs
field or the QCD condensates) do not contribute to the vacuum
entropy. In what concerns the zero-point fluctuations, they have,
properly speaking, an infinite entropy density, because (if we do
not impose any energy cutoff) the number of modes is infinite. But
if we regulate their energy, by introducing an ultraviolet cutoff
$m$, we also regulate their entropy. A simple estimation of the
resulting entropy bound can be derived as follows.

Limiting the energy-momentum space associated to the zero-point
fluctuations leads to the quantization of their configuration
space, with a minimum size given by $l \approx m^{-1}$. This
results in a superior bound to the number of available degrees of
freedom in a given volume, say, the volume inside the Hubble
horizon. The maximum number $N_{max}$ of observable degrees of
freedom will be of the order of $V/l^3$, where $V$ is the Hubble
volume. That is,
\begin{equation}\label{Nmax1}
\label{N} N_{max} \approx \left(\frac{R}{l}\right)^3 \approx
\left(\frac{m}{H}\right)^3.
\end{equation}

Note that this superior bound dominates over the matter entropy.
Indeed, taking for $H$ the value observed nowadays, $H \approx
70\; km.s^{-1}.Mpc^{-1}$, we obtain from (\ref{N}) (with $m$
equals to the pion mass, see below) $N_{max} \approx 10^{122}$, a
value that predominates over the matter entropy referred above, of
order $10^{101}$. Since $N_{max}$ increases with $R^3$, this
dominance will remain valid in the future, and we can consider the
entropy bound of the observable universe as given by (\ref{N}).

On the other hand, we have seen that this bound should be equal to
\begin{equation}
\label{Nmax} N \approx \left(\frac{R}{l_p}\right)^2 = (H
l_p)^{-2}.
\end{equation}
Therefore, we can identify (\ref{Nmax}) with (\ref{N}), which
leads to equation (3), that is, to the Eddington-Weinberg
relation.

But why has $m$ the order of the pion mass? The common belief is
that a natural cutoff for the vacuum fluctuations should be given
by the Planck mass $m_p=l_p^{-1}$, because at the Planck scale the
classical picture of spacetime breaks down. However, it is not
difficult to see that, equating (\ref{Nmax}) to (\ref{N}) with
$m=m_p$, one obtains a Hubble radius of the order of $l_p$, which
is not consistent with our universe.

One can also argue that the fluctuations of other fields than
quarks and gluons contribute to the vacuum entropy as well. It is
then intriguing that just the pion mass enters in the
Eddington-Weinberg relation. Nevertheless, let us remark that in a
curved spacetime the different sectors of the Standard Model of
particles interactions are coupled by gravity. Then one may expect
that (at late times) all the interacting vacuum fields tend to a
state of thermodynamic equilibrium, at a temperature equal or
below the temperature of the last vacuum phase transition. But the
last of such transitions was the chiral transition of QCD
\cite{Matthews}, at a temperature given by $\Lambda_{QCD}\approx
150$ MeV, of the order of the pion mass.

\section{Holography and decaying $\Lambda$ solutions}

As we have seen, the energy inside the Hubble horizon of the
spatially flat Friedmann universe is given by
\begin{equation}
E = \frac{R}{2G} = 4\pi \lambda R^2
\end{equation}
(where we have used (3)). Therefore, when the Hubble radius has an
increment $dR$ the energy content increases by
\begin{equation} \label{dE1}
dE = 8 \pi \lambda R\; dR = dR/G.
\end{equation}

On the other hand, in the case of a universe composed by dust and
a vacuum term with pressure $p = - \Lambda/(8 \pi G)$, as in
(5)-(6), the general thermodynamic expression for the energy
variation is given by
\begin{equation}
\label{dE2} dE = T dS - p\; dV = T d\left(\frac{A}{\alpha
G}\right) + \frac{\Lambda}{8\pi G}\; dV,
\end{equation}
where the holographic prescription $S = A/(\alpha G)$ has been
used (the positive dimensionless constant $\alpha$ is equal to $4$
in the black-hole case). Equating (\ref{dE2}) to (\ref{dE1}), and
using (3), $A = 4 \pi R^2$, and $V = 4\pi R^3/3$, we obtain
\begin{equation} \label{T}
T = \frac{\alpha H}{12\pi} \left( 1 - \frac{\Lambda}{2H^2}
\right).
\end{equation}

This last result has important implications in relation to the
cosmological solutions we have found in section II. First of all,
we see that (in the case of a non-negative cosmological constant)
$T\leq \alpha H/(12\pi)$. As far as the universe expands, the
temperature decreases, and we have two different possibilities.
The first one is that the universe tends asymptotically to a de
Sitter universe, with constant Hubble radius and constant
temperature (this corresponds to the late time solution with $n=0$
of section II). But in a de Sitter universe the Hubble constant is
given by $H=\sqrt{\Lambda/3}$. Therefore, from (\ref{T}) we have
$T = -\alpha H /(24\pi)$. Since $\alpha$ is a positive constant,
its clear that such a solution is not acceptable from a
thermodynamic point of view.

The second possibility is that the Hubble sphere expands forever,
with $R\rightarrow \infty$ (our late time solutions with positive
$n$). Leading (27) into (\ref{T}), we obtain
\begin{equation}
T = (n-1) \frac{\alpha H}{24\pi}.
\end{equation}
In this case, the temperature tends asymptotically to zero as far
as the universe tends to a flat spacetime. But the condition
$T\geq 0$ imposes $n\geq 1$. That is, the accelerating solutions
(including the case $n=0$, already discussed) are forbidden from
this thermodynamic viewpoint. On the other hand, the late time
coasting solution ($n=1$) corresponds to $T/H = 0$. In other
words, in the coasting expansion the temperature tends to zero
faster than in the decelerating cases.

If the above analysis is repeated in the context of a model with
varying $\Lambda$ but constant $G$, the same conclusion is
obtained: among the non-decelerating solutions, the coasting
expansion is the only acceptable from a holographic perspective.
However, in this case we would not be in accordance with the
observed relative matter density, as already discussed. Note also
that the holographic  expression (\ref{T}) suggests naturally the
$H^{2}$ dependence for $\Lambda$ discussed by several authors in
different contexts \cite{Ademir,Shapiro}.

\section{Conclusions}

The asymptotic scenario proposed here has several interesting
features, from both theoretical and observational viewpoints. A
coasting expansion seems to provide the best fit of the recent
observations of supernovas and radio sources \cite{Dev}.
Furthermore, it leads to a universe age given by $Ht=1$, once
again in good accordance with the observational limits \cite{age}.
Finally, a relative matter density equals to $1/3$ is in
remarkable agreement with observations \cite{omega}, specially if
we remember that it is an exact result in our context, and,
perhaps more important, it does not depend on any fine tuning of
model parameters or initial conditions.

From a theoretical perspective, the variation of the vacuum energy
density with the cosmic expansion is naturally expected on the
basis of quantum field theories in curved backgrounds
\cite{Ralf,Shapiro}, as well as the association of a temperature
to the vacuum of a curved spacetime. As discussed in the
Introduction, the same can be said about the variation of the
gravitational constant (and of the fine structure constant). The
empirical law initially used for the variation of $G$ (the
Eddington-Weinberg relation, expression of Dirac's large numbers
coincidence \cite{Ademir3,GS}) can be justified on the basis of
the holographic principle (see also \cite{GS,GRF}). Moreover, the
late time coasting expansion is, among the non-decelerating
solutions, the only compatible with a holographic, thermodynamic
analysis.

Probably, it is not necessary to say that our analysis cannot be
considered complete. First of all, the holographic idea we have
adopted to justify the Eddington-Weinberg relation is still just a
conjecture, despite the several theoretical developments this
subject has received in the last years \cite{Bousso}. On the other
hand, the Friedmann equation (1) applied to this varying $G$
context is something to be theoretically validated, for example in
the realm of scalar-tensor theories \cite{Bertolami,Pavon}. It is
possible to verify that our solutions, together with equation (1),
are solutions of Brans-Dicke theories, and an exhaustive study
about this will be shown elsewhere.

Finally, let us remind that our discussion is restricted to the
late eras of the expanding Universe. The study of a complete
scenario, including the hot phase at early times, depends on the
establishment of exact variation laws for $\Lambda$ and $G$, which
will depend on the development of a field theoretical treatment of
the problem in a curved, expanding background. Such a study should
also explain why the space is nearly flat, perhaps by means of a
fundamental justification of inflation which may ultimately be
related to a definite decaying law for $\Lambda$ at very short
times.

The authors are thankful to R. Abramo, A. I. Arbab, O. Bertolami,
M. V. John, T. Mongan, T. Padmanabhan, A. Saa, J. Sol\`a, J.-P.
Uzan, and R. G. Vishwakarma for useful discussions and
bibliographic suggestions. This work was partially supported by
CNPq.


\begin{thebibliography}{}

\bibitem{Starobinsky} V. Sahni and A. Starobinsky, Int. J. Mod. Phys. D {\bf 9}, 373
(2000); P. J. E. Peebles and B. Ratra, Rev. Mod. Phys. {\bf 75},
559 (2003).

\bibitem{OT} M. Ozer and O. Taha, Phys. Lett. A {\bf171}, 363
(1986); Nucl. Phys. B {\bf{287}}, 776 (1987).

\bibitem{Bertolami} O. Bertolami, Nuovo Cimento {\bf 93}, 36 (1986).

\bibitem{PR} P. J. E. Peebles and B. Ratra, Astrophys. J.  {\bf 325}, L17
(1988); B. Ratra and P. J. E. Peebles, Phys. Rev. D {\bf 37}, 3406
(1988).

\bibitem{Wu} W. Chen and Y.S. Wu, Phys. Rev. D {\bf 41}, 695 (1990).

\bibitem{Ademir} J. C. Carvalho, J. A. S. Lima and I. Waga, Phys. Rev. D {\bf
46}, 2404 (1992).

\bibitem{Ademir3} J. A. S. Lima and J. C. Carvalho, Gen. Rel. Grav.
{\bf 26}, 909 (1994).

\bibitem{Lima} J. A. S. Lima and J. M. F. Maia, Phys. Rev. D {\bf 49},
5597(1994); J. A. S. Lima and M. Trodden, Phys. Rev. D {\bf 53},
4280 (1996); J. A. S. Lima, J. M. F. Maia and N. Pires, IAU
Simposium {\bf 198}, 111 (2000); J. V. Cunha, J. A. S. Lima and N.
Pires, Astron. and Astrophys. {\bf 390}, 809 (2002).

\bibitem{John} M. V. John and K. B. Joseph, Phys. Rev. D {\bf 61}, 087304
(2000); M. Novello, J. Barcelos-Neto and J. M. Salim, Class.
Quant. Grav. {\bf 18}, 1261 (2001); {\bf 19}, 3107 (2002); R. G.
Vishwakarma, Class. Quant. Grav. {\bf 18}, 1159 (2001); {\bf 19},
4747 (2002); J. S. Alcaniz and J. M. F. Maia, Phys. Rev. D {\bf
67}, 043502 (2003); A. I. Arbab, Class. Quant. Grav. {\bf 20}, 93
(2003).

\bibitem{MGX} S. Carneiro, gr-qc/0307114, Tenth Marcel Grossmann
Meeting, Rio de Janeiro, 2003.

\bibitem{GRF} S. Carneiro, Int. J. Mod. Phys. D {\bf 12}, 1669 (2003);
gr-qc/0206064.

\bibitem{Aldrovandi}R. Aldrovandi, J. P. Beltr\'an and J. G.
Pereira, gr-qc/0312017.

\bibitem{Ralf} R. Sch\"utzhold, Phys. Rev. Lett. {\bf 89}, 081302
(2002); Int. J. Mod. Phys. A {\bf 17}, 4359 (2002).

\bibitem{Shapiro} I. L. Shapiro and J. Sol\`a, astro-ph/0401015.

\bibitem{Ranada} A. F. Ra\~nada, Europhys. Lett. {\bf 61}, 174
(2003).

\bibitem{Alfa} J. K. Webb {\it et al.}, Phys. Rev. Lett. {\bf 87}, 091301 (2001).

\bibitem{Dirac} P. A. M. Dirac, Nature {\bf 139}, 323 (1937); Proc.
Roy. Soc. Lond. {\bf 165}, 199 (1938).

\bibitem{frances} J.-P. Uzan, Rev. Mod. Phys. {\bf 75}, 403 (2003).

\bibitem{GS} G. A. Mena Marug\'{a}n and S. Carneiro, Phys. Rev. D {\bf 65},
087303 (2002).

\bibitem{Bousso} R. Bousso, Rev. Mod. Phys. {\bf 74}, 825 (2002); T. Jacobson and
R. Parentani, Found. Phys. {\bf 33}, 323 (2003); T. Padmanabhan,
gr-qc/0311036.

\bibitem{geral} By using (8), (11) and (18), it is possible to
rewrite the solution (17) as $a=a_0[1+nH_0(t-t_0)]^{1/n}$, where
$a_0$ and $H_0$ are, respectively, the scale factor and the Hubble
parameter at $t=t_0$. In the limit $n \rightarrow 0$, we obtain
the de Sitter solution $a=a_0\exp[H_0(t-t_0)]$, as should be.

\bibitem{Dev} A. Dev, M. Sethi and D. Lohiya, Phys. Lett. B {\bf 504}, 207
(2001); D. Jair, A. Dev and J. S. Alcaniz, Class. Quant. Grav.
{\bf 20}, 4163 (2003). See also M. V. John and J. V. Narlikar,
Phys. Rev. D {\bf 65}, 043506 (2002).

\bibitem{age} B. M. S. Hansen {\it et al.}, Astrophys. J.
{\bf 574}, L155 (2002).

\bibitem{omega} M. Tegmark {\it et al.}, Phys. Rev. D {\bf 69}, 103501
(2004).

\bibitem{Zeldovich} In Planck units, equations
(26)-(27) lead to $\Lambda \approx m^6$ (except for $n=3$, in
which case the cosmological constant is zero). This result is just
the empirical relation originally considered by Zel'dovich, 37
years ago \cite{Ademir3,GRF}.

\bibitem{Wesson} For $R = H^{-1} \approx \Lambda^{-1/2}$, we
have $m_0 \approx \sqrt{\Lambda}$. Paul Wesson (Mod. Phys. Lett. A
{\bf 19}, 1995 (2004)) has recently proposed such a quantum of
mass, on the basis of dimensional analysis and high dimensional
theories.

\bibitem{Penrose} R. Penrose, {\it The Emperor's New Mind} (Oxford
University Press, 1989), chapter $7$.

\bibitem{Matthews} R. A. J. Matthews, Astron. \& Geophys. {\bf 39}, 19 (1998).

\bibitem{Pavon} N. Banerjee and D. Pav\'on, Class. Quant. Grav. {\bf 18}, 593
(2001); Phys. Rev. D {\bf 63}, 043504 (2001).

\end{thebibliography}
\end{document}